\documentclass[prb,
reprint,
 amsmath,amssymb,
 aps,
]{revtex4-2}

\usepackage{graphicx}
\usepackage{dcolumn}
\usepackage{bm}

\newcommand{\wv}[1]{{#1}}

\usepackage{physics,amsmath,amssymb,xcolor}

\begin{document}

\preprint{APS/123-QED}

\title{Optical resonators constitute a universal spin simulator}

\author{Wouter Verstraelen}
\email{verstraelenwouter@gmail.com}
\affiliation{Majulab, International Joint Research Unit UMI 3654, CNRS, Université Côte d’Azur, Sorbonne Université, National University of Singapore, Nanyang Technological University, Singapore, Singapore 117543}
\affiliation{Division of Physics and Applied Physics, School of Physical and Mathematical Sciences, Nanyang Technological University, Singapore 637371, Singapore}

\author{Timothy C.H. Liew}
\email{tchliew@gmail.com}
\affiliation{Majulab, International Joint Research Unit UMI 3654, CNRS, Université Côte d’Azur, Sorbonne Université, National University of Singapore, Nanyang Technological University, Singapore, Singapore 117543}
\affiliation{Division of Physics and Applied Physics, School of Physical and Mathematical Sciences, Nanyang Technological University, Singapore 637371, Singapore}

\date{\today}

\begin{abstract}
NP-hard computational problems can be efficiently recast as finding the ground state of an effective spin model. However, to date no convenient setup exists that can universally simulate all of them, even for a fixed problem size. Here we present such a setup, \wv{the \emph{universal homogeneous spin simularor (UHSS)}}  using a series of optical (or polaritonic) resonators arranged in a chain using the geometry \wv{based on the one} introduced in [Phys. Rev. Applied 21, 024057 (2024)]. We demonstrate by example how the simulator solves Hamiltonian Cycle and traveling salesman problems, and show that it generalises to any NP-hard problem of arbitrary size. \wv{This approach works because it allows arbitrary long-range interactions in the spin model to be mapped on an optical system with only nearest-neighbor physical interactions.}
\end{abstract}

\maketitle

\section{Introduction}

Finding the ground states of spin models can be NP-hard \cite{Stein_Newman_2013}. That means that through obtaining these, one can solve all kinds of complex computational decision and optimisation problems \cite{Lucas14}. 
Optics and polaritonics -- the latter using photons hybridized with matter excitations to enhance nonlinearities \cite{Sanvitto2016} -- can provide convenient platforms to simulate these behaviors for aforementioned computational tasks \cite{Stroev2023,Opala23,Kavokin2022,Mohseni2022} as well as achieve fundamental insights \cite{Angelakis_2017,Amo2016,Verstraelen2020,Marsh2024}, with growing progress for implementation using plain laser light \cite{Nixon2013,Yamamoto2017,Pierangeli2019}, polaritons~\cite{Ohadi2015,Ohadi2017,Berloff2017,Kalinin2020,Kyriienko2019,Sigurdsson2019}, or photon condensates~\cite{Toebes2022,Mattschas2024}. Typically, the mapping becomes inaccurate by heterogeneity of the amplitudes \cite{Strinati2021PRL,Cummins2023}, however this issue can be avoided exactly \cite{Verstraelen24} or managed approximately \cite{Shi24,Leleu2019}.

Despite all this progress on optical spin simulators that may work in principle, they tend to suffer from one major limitation that holds them back from solving many complex problems on a larger scale: their lack of flexibility. Typically, one fixed hardware setup would only solve one specific instance of a problem: rarely would it be worthwile to build a simulator for just a single task.
Some approaches (such as spatial light modulators) that are adaptive may offer hope to simulating different configurations with more flexibility. Still, such approaches tend to be restricted in the graphs that can be implemented, and in particular require the graph to be planar with nearest-neighbour couplings.

To be able to solve many instances of a computational problem, one needs the ability to use a fixed setup for all of these. Such a setup is then a `universal' simulator \cite{Feynman1982}. Formally, an Ising model on a sufficiently large regular square grid is such a universal simulator \cite{Cuevas16}. For many NP-hard problems, it is known how to recast them as Quadratic Unconstrained Binary Optimisation (QUBO) problems, that are in turn closely connected to Ising graphs.
However, the mapping of a generic Ising graph to an Ising model on a square lattice, according to this approach \cite{Cuevas16}, would introduce a massive overhead in the problem size. For example, the mapping of a generic graph to a planar graph itself is an intermediate necessary step, and already needs the introduction of a `crossing gadget' of 22 additional vertices and 40 additional edges \cite{Cuevas16} for each avoided crossing.

We thus wonder if there is hope for a universal simulator that is reasonably scalable and accessible. \wv{The use of feedback may be helpful in this regard \cite{McMahon2016,Byrnes_2011}.} In this work, we will show how to achieve \wv{the aforementioned task} by a single Ising spin chain in 1D, subject -in principle- to feedback induced long-range interactions. As we show, this model can be straightforwardly implemented in the manner of \cite{Verstraelen24} in a fixed setup with optical resonators such as micropillars, cavities or waveguides\wv{, which we call the UHSS}. In the next section \ref{sec:Setup}, we introduce the \wv{UHSS}, and demonstrate abilities to solve Ising spin problems in \ref{sec:spinglassproblem}. We then show show how to solve a few NP-hard problems of practical importance in section \ref{sec:NPhardproblems} and finally conclude in section \ref{sec:conclusions}.

\section{Setup \label{sec:Setup}}

\subsection{Ising spin graph}

Recall that a general (classical) Ising Hamiltonian is given by 
\begin{equation}\label{eq:Isingwithoffset}
    E_\text{spin}(s_1,\ldots,s_N)=-\sum_{i=1,j>i}^NJ_{ij}s_i s_j -\sum_{i=1}^N h_i s_i,
\end{equation}
where $s_i=\pm1$\wv{, $J_{ij}$ represents the interaction strengths and $h_i$ external fields.}
We can in fact rewrite this as
\begin{equation}
    E_\text{spin}(s_1,\ldots,s_N)=-\sum_{i=0,j>i}^NJ_{ij}s_i s_j \label{eq:IsingHwithoutouffset}
\end{equation}
provided we set $s_0=1$ (which can always be achieved by a global rotation) and $J_{i0}=h_i$, thus introducing an $N+1$th spin.

We now study this problem in a polaritonic spin simulator, composed of a graph of spatially separated polariton condensates described by macroscopic wavefunctions $\psi_i$ ($i$ being a site index). \wv{Thanks to growing experimental progress, such structures are increasingly available, even at room temperature \cite{Ghosh22}}.

As in e.g. Ref. \cite{Berloff2017}, each spin $s_i$ in a considered graph is mapped to the phase of $\psi_i$, respectively. Rather than have $\psi_i$ directly coupled, we follow Ref. \cite{Verstraelen24} where it was proposed that the coupling $J_{ij}$ in the spin graph between $s_i$ and $s_j$ could be realized with pairs of additional polariton condensates (or optical resonators) described with wavefunctions $\chi_{ij}^S$ and $\chi_{ij}^A$. Here $\chi_{ij}^S$ is linearly coupled with positive coupling constant to both $\psi_i$ and $\psi_j$ (`symmetrically'), while $\chi_{ij}^A$ is coupled with positive and negative coupling constants to $\psi_i$ and $\psi_j$ (`antisymmetrically'), respectively (See Fig. \ref{fig:twospinunit}) In a steady state, the result has by symmetry:

\begin{figure}
    \centering    
    \includegraphics[width=0.7\linewidth]{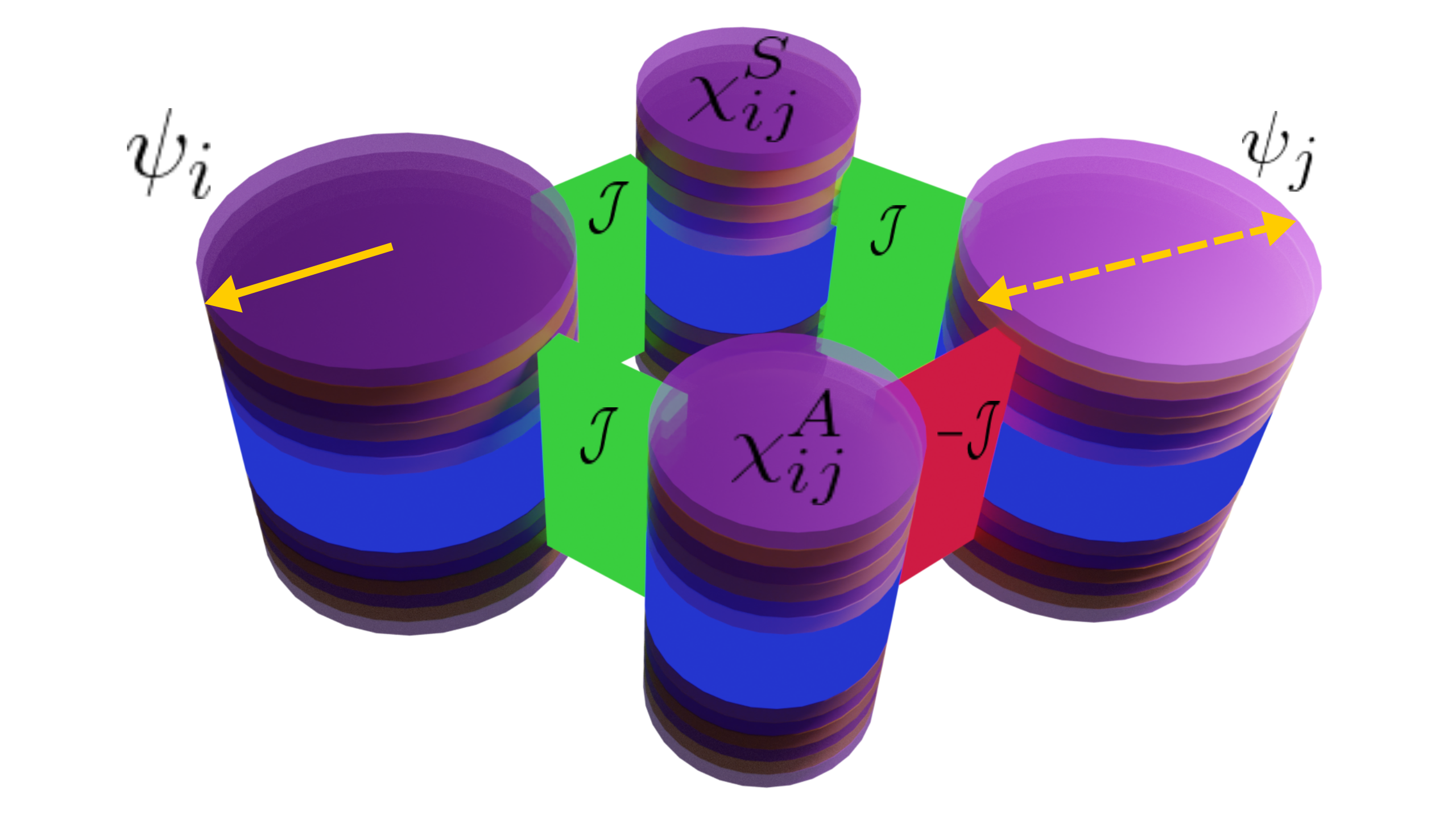}
    \caption{The unit coupling two effective spin wavefunctions $\psi_i,\,\psi_j$, here depicted with micropillars. The phase of $\psi_j$ is either the same or opposite to $\psi_i$, and depending on this, $\chi^S_{ij}$ or $\chi^A_{ij}$ will have have finite occupation, respectively.}
    \label{fig:twospinunit}
\end{figure}

\begin{equation}
    \chi^{S/A}_{ij}\propto \psi_i \pm \psi_j, 
\end{equation}

such that
\begin{equation}
\abs{\chi^{S/A}_{ij}}^2 \propto (\abs{\psi_i}^2+\abs{\psi_j}^2\pm (\psi_i^*\psi_j+\psi_j^*\psi_i)).
\end{equation}

Note that, this scheme avoids amplitude heterogeneity, $\abs{\psi_i}=\abs{\psi_j}:=\abs{\psi}$, and in presence of a nonlinear loss process to the $\chi$-modes, $\psi_i=\pm \psi_j$ (the proofs are given in Ref.\cite{Verstraelen24}). We can now define $s_i=\psi_i/\psi_0=\pm 1$ and obtain

\begin{equation}
\abs{\chi^{S/A}_{ij}}^2 \propto (1\pm s_i s_j).   
\end{equation}

Note that $s_0=1$ is fixed by this definition automatically. Rewriting this expression, we obtain

\begin{align}\label{eq:intensitiestospins}
\frac{|\chi_{ij}^{S}|^2-|\chi_{ij}^{A}|^2}{|\chi_{ij}^{S}|^2+|\chi_{ij}^{A}|^2}
=s_i s_j.
\end{align}

Filling in \eqref{eq:IsingHwithoutouffset}

\begin{equation}\label{eq:Expinfromchis}
    E_\text{spin}(\ldots\abs{\chi_{ij}}^2\ldots)=-\sum_{i=0,j>i}^NJ_{ij} \frac{|\chi_{ij}^{S}|^2-|\chi_{ij}^{A}|^2}{|\chi_{ij}^{S}|^2+|\chi_{ij}^{A}|^2}
\end{equation}

We have thus completely represented $E_\text{spin}$ for arbitrary Ising models in the polariton degrees of freedom. 

Upon condensation, the phases of $\psi_i$ are in principle chosen randomly and thus sample a possible effective spin configuration. It is possible to bias the system into condensing into specific states (such as the ground state) by applying a feedback mechanism that uses the measurable $E_\text{spin}$ as an effective energy (see below).
Note that the ratios on the right-hand side of (\ref{eq:Expinfromchis}) represent the relative population imbalance (restricted to $\pm 1$), with denominators that are constant across configurations.

\subsection{Universal simulator on a chain}

In the previous subsection, we have seen how any fixed Ising spin graph can be mapped to an optical setup that mimics its geometry. The fixed setup can thus only solve a single instance of a problem, making it quite restrictive. We are interested in a universal optical simulator with fixed geometry, that can solve arbitrary spin graph problems. In principle, some models such as nearest-neighbour Ising models on a square lattice \cite{Cuevas16} could form such a universal geometry, but this would introduce a massive overhead, limiting usefulness.

Here, instead, we show how we can achieve the universal simulation in a linear chain, \wv{the UHSS,} taking into consideration the possibility of effective longer range interactions properly.  Such interactions can be implemented by modification of the feedback scheme rather than requiring any physical mechanism of long-range inter-particle interaction.

Starting from \eqref{eq:IsingHwithoutouffset} and noting that $s_i^2=1$, we obtain

\begin{align}
    E_\text{spin}(s_1,\ldots,s_N)&=-\sum_{i=0,j>i}^N J_{ij}s_i s_j\\
    &=-\sum_{i=0,j>i}^N J_{ij}s_i\left(\prod_{k=i+1}^{j-1} s_k^2\right)  s_j\\
    &=-\sum_{i=0,j>i}^N J_{ij} \prod_{k=i}^{j-1} s_k s_{k+1},
\end{align}

thus $E_{spin}$ can be represented entirely as a polynomial in nearest neighbour couplings in a one dimensional chain of $N+1$ sites. Finally, to map to the optical simulator, we use \eqref{eq:intensitiestospins} and obtain 

\begin{equation}\label{eq:Spinefromoccupsinchain}
    E_\text{spin}=-\sum_{i=0,j>i}^N J_{ij} \prod_{k=i}^{j-1} \frac{|\chi_{k,k+1}^{S}|^2-|\chi_{k,k+1}^{A}|^2}{|\chi_{k,k+1}^{S}|^2+|\chi_{k,k+1}^{A}|^2}. 
\end{equation}

\section{The spin glass problem as an NP-hard problem \label{sec:spinglassproblem}}
Before turning to a class of more specific useful NP-hard problems itself, we start with the spin-model physics itself.  

As a brief recap, if an Ising system is known to not be frustrated, meaning that the ground state minimizes all terms of \eqref{eq:Isingwithoffset} at once, this ground state is easily found by separately minimising these terms. This will take only \emph{Polynomial time} in system size and thus belongs to the \emph{P} complexity class.
For generic frustrated spin models,  such a construction is impossible, and the problem of finding ground states takes more than polynomial time, it is known in fact to belong to the \emph{NP-hard} class. As one can easily verify $E_\text{spin}$ given a configuration $\{s_i\}$ in polynomial time using \eqref{eq:Isingwithoffset}, the \emph{decision problem} version (is there a configuration with $E_\text{spin}\leq E_\text{spin}^0$?) is by definition an \emph{NP}-problem itself. Being simultaneously \emph{NP-hard} and \emph{NP}, it is \emph{NP-complete} \cite{Lucas14}.

\subsection{Existance of solutions of certain $E_\text{spin}$}

Starting with the decision version of the above problem, we look for the existence of solutions at specific values of $E_\textbf{spin}$.

Following Ref.~\cite{Verstraelen24}, the \wv{mean-field} evolution equations for both the spin resonatators $\psi_j$ and the the coupling resonators $\chi_{j,j+1}^{S/A}$ are given by

\begin{align}\label{eq:GPE}
i\frac{\partial\psi_j}{\partial t}&=i\left(\frac{P_j(t)-\gamma}{2}-\Gamma_\mathrm{NL}|\psi_j|^2\right)\psi_j\nonumber\\&\hspace{1.5cm}-\mathcal{J}\left(\chi_{j,j-1}^S-\chi_{j,j-1}^A+\chi_{j,j+1}^S+\chi_{j,j+1}^A\right)\nonumber\\
i\frac{\partial\chi_{j,j+1}^{S}}{\partial t}&=i\left(\frac{-\gamma}{2}-\Gamma_\text{NL}\abs{\chi_{j,j+1}^{S}}^2\right)\chi_{j,j+1}^{S}
-\mathcal{J}\left(\psi_j+\psi_{j+1}\right)\nonumber\\
i\frac{\partial\chi_{j,j+1}^{A}}{\partial t}&=i\left(\frac{-\gamma}{2}-\Gamma_\text{NL}\abs{\chi_{j,j+1}^{A}}^2\right)\chi_{j,j+1}^{A}
-\mathcal{J}\left(\psi_j-\psi_{j+1}\right)
\end{align}
where $\gamma$ and $\Gamma_\text{NL}$ are the linear and nonlinear loss processes and $\mathcal{J}$ the hopping amplitude between modes. One finds that the net linear loss to the spin sites $\psi_j$ at zero intensity is $\gamma'_j=\gamma+(2-\delta_{j,1}-\delta_{j,N})\frac{ 8\mathcal{J}^2}{\gamma}$ (the latter term representing losses through adjecent coupling sites). This means that one has a finite and uniform occupation in the steady state for $P_j=\gamma'_j+P_0,\, P_0>0$ \footnote{To ensure complete homogeneity at finite occupation in the steady state, $\gamma'_j=\gamma+(2-\delta_{j,1}-\delta_{j,N})\frac{ 8\mathcal{J}^2}{\gamma+2\abs{\chi}^2\Gamma_{NL}}$, where $\abs{\chi}^2$ can be obtained self-consistently, or alternatively extra dangling sites can be added \cite{Verstraelen24}}.
One can then extract $E_\text{spin}$ using \eqref{eq:Spinefromoccupsinchain}.  

\wv{The use of driven-dissipative Gross-pitaevskii equations as  \eqref{eq:GPE} is well known to be valid for the description of polariton condensates \cite{Carusotto2013} and related systems \cite{Aranson2002} well above optical threshold where there is a nonzero mean-field. It is complemented by a the addition of a small random perturbation to the initial state to account for fluctuations.}

Based on this extracted value of $E_\text{spin}$, we can either accept or reject the configuration, by using a "feedback" \wv{stage} where 
$\wv{P_j=} P^f_j\wv{:}=\gamma'_j+f(E_\text{spin})$ and \wv{the function} $f$ is defined such that $f(E_\text{spin})>0$ for the desired values of $E_\text{spin}$ only. For example in the case of the question\wv{, for any arbitrary value $E_\text{spin}^0$,} whether $E_\text{spin}\leq E_\text{spin}^0$ exist, one can use 

$$f(E_\text{spin})=-\mu(E_\text{spin}-E_\text{spin}^0)$$ 

\wv{for a positive constant $\mu$.}

Working in a pulsed regime, the simulator samples many configurations at a time as an effective Monte Carlo solver. 
\wv{If during a given pulse $f(E_\text{spin})>0$, the mean-field amplitudes $\psi_j$ will retain a finite value, keep their relative phases and thus the current spin configuration. In the opposite case $f(E_\text{spin})<0$, the amplitudes $\psi_j$ will decay back to the vacuum, so that the formation of a new spin configuration at the beginning of the next pulse will be dominated by the noise. The physical time required for each pulse will be strongly platform-dependent and would typically be a few tens of times the lifetime $\gamma^{-1}$ in a polariton system. Experimentally, a polariton buildup time 200~ps has been observed in GaAs \cite{Anton2014}, as well as a switching time of 360~fs and a response time of 80~fs in ZnO \cite{Chen2022,Li2024}.
} 

If\wv{, after many pulses,} the simulator settles in a steady state, it means that an accepted value $E_\text{spin}\leq E_\text{spin}^0$ was found. If, on the other hand, no steady state is found after sufficient evolution, it indicates the absence of such a solution. \wv{Note that only at most one spin solution will be found at each run of the simulator, even in the case of degeneracy. This is likely sufficient for practical purposes, if all solutions need to be found, one can run the aforementioned procedure several times.}

\subsection{optimising for the ground state\label{subsec:GSoptim}}

We can now move to the optimisation version of the problem, finding the ground state of the Ising model. An approach to this aim is to repeat the decision problem study outlined above for many values of $E_\text{spin}^0$ differing by a value $\varepsilon$. The ground state, up to precision $\varepsilon$, would then be found at the lowest value of $E_\text{spin}^0$ where a convergence to a steady state happens.

In practice, it can be more convenient to increase the value of $E_\text{spin}^0$ continuously across all pulses within a range $\left[E_\text{spin}^{0\,\text{(min)}}, E_\text{spin}^{0\,\text{(max)}}\right]$. The precision is then only subject to the sufficient amount of pulses to sample all phase configurations for each energy well. 

The simplest scheme has a linear increase of the form
\begin{align}
E_\text{spin}^{0\,\text{(pulse)}}=&\left(1-\frac{\text{pulse}}{\text{amount of pulses}}\right)E_\text{spin}^{0\,\text{(min)}}\nonumber\\ &+\left(\frac{\text{pulse}}{\text{amount of pulses}}\right)E_\text{spin}^{0\,\text{(max)}}    
\end{align}

\begin{figure}
    \centering
    \includegraphics[width=0.7\linewidth]{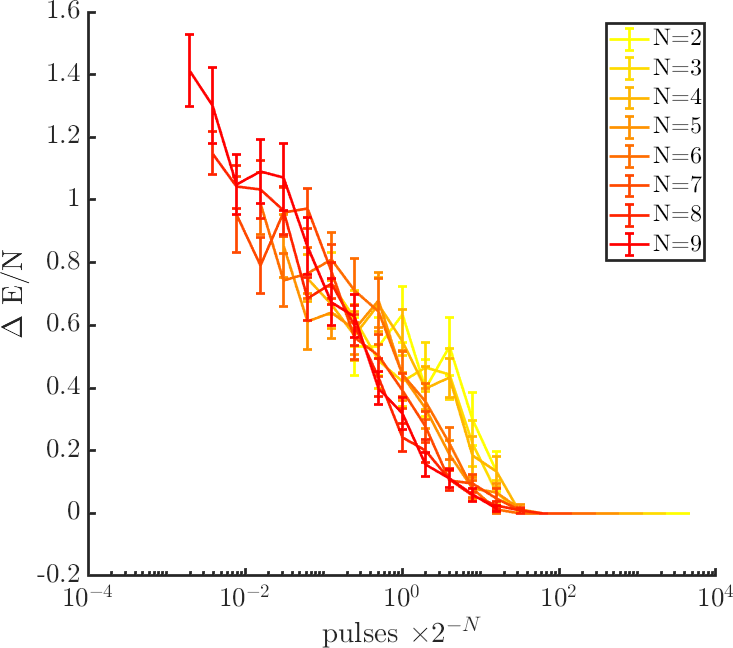}
    \caption{Rescaled excess energy about ground state for graphs implemented on a chain of different lengths. We see that the ground state is always reached with enough iterations, and the required amount of iterations scales in fact better than $\mathcal{O}(2^N)$. \wv{The required duration of each pulse can be estimated as a few tens of polariton lifetimes. Parameters used as in \cite{Verstraelen24}}.}
    \label{fig:scaling}
\end{figure}

In Fig.\ref{fig:scaling}, we  numerically show the workings of this scheme for different system sizes, and using $E_\text{spin}^{0\,\text{(min)}}=-50; E_\text{spin}^{0\,\text{(max)}}=0$. 
    We observe that with a sufficient amount of iterations, the ground state is always found. The amount of iterations required through this scheme grows in fact slower than $2^N$. \wv{The workings of this scheme are independent of the energy landscape (which will differ between frustrated and non-frustrated systems), although the precision will be limited by the extend by which the spectral gaps in $E_\text{spin}$ can be resolved.}

\section{Solving Practical NP-hard problems \label{sec:NPhardproblems}}

Above, it has has been shown that the chain configuration can universally simulate all Ising spin graphs and obtain their low energy configurations. Since the latter problem is NP-hard, it is possible to solve all other such NP-hard problems with only polynomial overhead. For many of these problems, the mapping to an Ising graph is explicitly known \cite{Lucas14}, making the chain simulator a convenient tool to solve them. Below, we demonstrate this by the explicit example of a Hamiltonian Cycle and Traveling Salesman problem, decision and optimisation problems.

\begin{figure}
    \centering
    \includegraphics[width=0.8\linewidth]{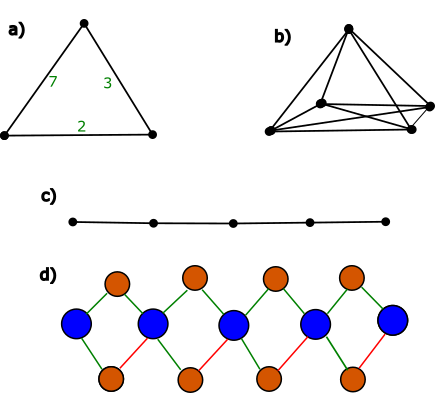}
    \caption{Overview of the different mappings. a) An NP problem is posed on a graph with $M$ nodes, in this case Hamiltonian Cycles or Traveling Salesman with $M=3$, and in the latter case weights attributed to each of the edges. b) It is represented on an Ising graph of $N=(M-1)^2+1$ nodes with a complex, potentially all-to-all connected, geometry. c) Our universal solver is able to represent the whole setup on a linear chain of effective spins. d) The full setup of the optical simulator, blue circles denote the $\psi$-resonators that represent the actual spins, whereas the orange  circles on the top and bottom denote the $\chi^S$ and $\chi^A$ coupling resonators respectively. Green(red) lines denote positive(negative) Josephson coupling}
    \label{fig:mappings}
\end{figure}

\subsection{Hamiltonian cycles problem}

The Hamiltonian Cycle problem, defined on a graph G with M nodes, is the decision problem that ask the following question:
``Does there exists a cyclic path along the edges, visiting all nodes exactly once''. We follow the mapping to an Ising problem of Ref \cite{Lucas14}, which requires a binary variable $x_i\in\{0,1\}$ for each combination of a node $v$ and a timestep $j$.  
One defines the effective `Hamiltonian' cost function

\begin{align}\label{eq:Hamcyclesx}
H_{HC}=&A\sum_{v=1}^n\left(1-\sum_{j=1}^N x_{v,j}\right)^2+A\sum_{j=1}^n\left(1-\sum_{v=1}^N x_{v,j}\right)^2\nonumber\\&+A\sum_{(uv)\not\in E}\sum_{j=1}^N x_{u,j} x_{v,j+1}   
\end{align}
with $A$ an arbitrary positive constant. The first two terms for \eqref{eq:Hamcyclesx} ensure that no node is visited twice and that only one node is visited at each timestep. The last term ensures that in subsequent timesteps, only adjacent sites can be visited.

\begin{figure}
    \centering
    \includegraphics[width=0.5\linewidth]{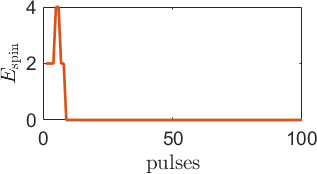}
    \caption{Simulation of a Hamiltonian cycles problem for the graph in Fig. \ref{fig:mappings}a. After a few pulses, the simulator reaches a solution with $E_\text{spin}=0$ and converges, demonstrating that the Hamiltonian cycle exists. Obtained through numerical simulation of the generalised Gross-Pitaevskii equations (\ref{eq:GPE}) for parameters: $A=100,\, \mu=1,\, \mathcal{J}/\Gamma_{NL}=0.5,\,\gamma/\Gamma_{NL}=4,\,P_0/\Gamma_{NL}=12$. Time interval per pulse: $1000\Gamma_{NL}^{-1}$, of which half is spent in the feedback and half in the readout stage.
}
    \label{fig:Hamcycle}
\end{figure}

The answer to the Hamiltonian cycle problem, the existance of such a cycle, is yes if and only if \eqref{eq:Hamcyclesx} has a ground state of energy $H_\text{HC}=0$.
To solve this problem on the universal spin simulator, we simply require a chain of length $N=(M-1)^2+1$ \footnote{The amount of variables $x$ or $s$ is reduced from the original $M^2$ assuming without loss of generality that the cycle starts in node 0. Additionally, this node node 0 also takes the role of $s_0$ in Eq. \eqref{eq:IsingHwithoutouffset}.} (Fig. \ref{fig:mappings}), and denote the spin sites by a single index $k=(v-1)(M-1)+j$. We can use $H_{HC}$ immediately for $E_\text{spin}$. To this aim, one
substitutes $x_i=\frac{s_i+1}{2}=\frac{s_i+s_0}{2}$ in \eqref{eq:Hamcyclesx}, set diagonal terms $s_i^2=1$ and substitute the off-diagonal $s_i s_j$ for the ratio of intermediate site intensities as in \eqref{eq:Spinefromoccupsinchain}. In Figure \ref{fig:Hamcycle}, we show numericaly how the existence of a cycle is found after several pulses.

\subsection{Traveling salesman decision problem}

\begin{figure}
    \centering
    \includegraphics[width=\linewidth]{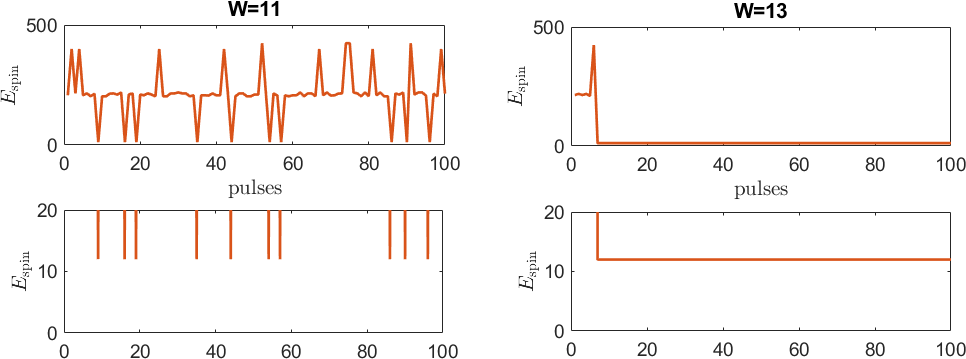}
    \caption{Traveling salesman problem, decision version for the graph in Fig. \ref{fig:mappings}a. No solution is found with $W<11$, but one is found with $W<13$. Parameters as in \ref{fig:Hamcycle}, with additionally $B=1$.}
    \label{fig:TSPdecision}
\end{figure}

The traveling salesman problem is an extension of the Hamiltonian cycles problem, where each edge of the graph has a weight $W_{uv}$ associated to it, and the question is not just whether a cyclic path exists, but if a cyclic \wv{path} of a total weight $\leq W$ exists. Its Hamiltonian is given by 
\eqref{eq:Hamcyclesx}

\begin{align}\label{eq:HamTSPx}
    H_\text{TSP}&=H_\text{HC}+B \sum_{(uv)\in E} W_{uv}\sum_{j=1}^N x_{u,j} x_{v,j+1}-B \,W \nonumber\\
    &=E_\text{spin}-E_\text{spin}^{(0)}
\end{align}
with $E_\text{spin}^{(0)}=B W$. And $A\gg B$ so that it is never favourable to violate the constraints set by $H_\text{HC}$
From there, the implementation on a chain proceeds as before. In Figure \ref{fig:TSPdecision}, we numerically study the traveling salesman decision problem on the same graph but now carrying weights (values depicted on Fig.\ref{fig:mappings} a). We see that a solution for $W=11$ is not found, but it is for $W=13$. \wv{We note that the Ising model in this example is in fact frustrated as the weights are all positive}.

\subsection{Traveling Salesman Optimisation problem}

\begin{figure}
    \centering
    \includegraphics[width=0.7\linewidth]{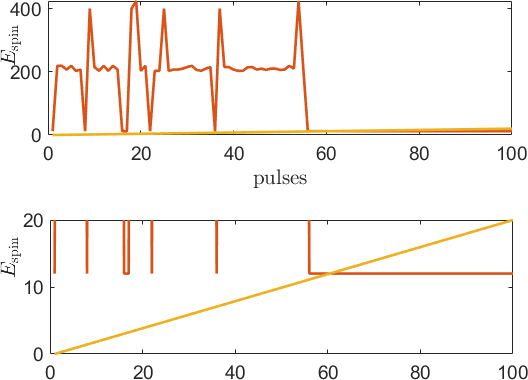}
    \caption{Traveling Salesman problem, optimisation version for the graph in Fig. \ref{fig:mappings}a. By continuously increasing $W$, the lowest energy solution to the problem is found, and therefore the path of lowest weight. Parameters as in the previous figures}
    \label{fig:TSPoptimisation}
\end{figure}

In the NP-hard optimisation version of the traveling salesman problem, probably the most known formulation, we seek the path with minimal \wv{weight ('}length\wv{')} $W$. This can be achieved from the same Hamiltonian \eqref{eq:HamTSPx}. However, now $W$ is no longer fixed, and across the many pulses, it increases slowly from $W^\text{(min)}$ to $W^\text{(max)}$, until the state converges for a certain value of $W^\text{(pulse)}$, in analogy with the spin ground state problem in \ref{subsec:GSoptim}. In Fig. \ref{fig:TSPoptimisation}, it is seen that the lowest path indeed found at $W=12$.

\section{Conclusions \label{sec:conclusions}}

Analogue spin simulation with light has drawn much experimental and theoretical interest, leading to a steady advance. It is largely directed at the prospects of solving NP-hard and NP-complete problems that are complicated in traditional platforms. 

This work takes a substantial further step in this direction. We have shown that a simple, non-tailored, geometry -an effective linear chain-, constitutes a \emph{universal} simulator, capable of solving any NP-hard task\wv{, the UHSS}. Building upon the previous work \cite{Verstraelen24}, the particular implementation with optical or polaritonic resonators keeps the amplitudes completely homogeneous and thus avoids the bias that hindered many earlier simulators. In fact, we have found from a scaling analysis of finding spin ground states that the correct solution will indeed be found exactly, provided that a sufficient amount of pulses is used. With a refined feedback process we have precise access to the effective spin energy, allowing the solution of decision problems as well as optimisation problems. 

\wv{The current work has focused on finding the ground state practically for individual frustrated spin graphs as they may arise from mappings of instances of NP-hard problems. On the other hand, one also studies the statistical aspects of such frustraded spin models, giving rich fundamental physics in terms of glass-like behaviour and replica-symmetry breaking \cite{Parisi2023}. There are important connections between these two research fields: for example, in simulated annealing one seeks the ground state by reducing the temperature to zero, however due to the complex energy landscape of a glass, this must be done exponentially slow \cite{Kirkpatrick83}. Many analogue simulators would similarly involve a mechanism to slowly reduce the amount of fluctuations relative to mean-field, and could thus be understood in terms of an effective temperature \cite{Verstraelen2020,Marsh2024,Yamamura2024}. The simulator presented here can be interpreted as having a disordered phase at effective temperature $T=\infty$ while it keeps looking for solutions and a $T=0$ phase after a solution has been found, and propagates to the subsequent pulses. It would be a an interesting open question if a refinement of this procedure could provide further enhancements. This relates to simulations further away from the mean-field limit, where time-dependent noise must be added to \eqref{eq:GPE}.} 

\wv{Slightly less trivial limitations of the simulator presented here relate to the fact that $E_\text{spin}$ must be extracted at every timestep from \eqref{eq:Spinefromoccupsinchain} with sufficient resolution, and $P_f$ adapted to it. While this overhead can be expected to be constant up to a certain limit. When for example the size of the bit registers becomes a limiting factor, eventual further scaling of these can be expected to be polynomial at most, since it constitutes a simple arithmetic task. and thus less than the exponential scaling of the amount of pulses.It is also independent of whether the underlying spin model is frustrated or not.
}

We have shown by explicit example how \wv{the scheme} works for Hamiltonian Cycle and Traveling Salesman problems. However, the same recipe would work for any NP-hard problem of interest. The only requirement is that its formulation as an Ising or QUBO problem is known. The setup thus provides a feasible pathway to achieve this task. \wv{It is universal and exact, likely at the cost of a slower execution time for very large systems compared to state-of-the-art solvers such as coherent Ising machines. This also provides the opportunity to seek further improvements to speed up this approach. This could involve also an extension to higher order spin models \cite{Bybee2023}}.
\wv{Another interesting question is} to which effect quantum effects could lead to further improvement \cite{Altman2021}, and how this would compare with alternative approaches for QUBO solving such as variational quantum algorithms on a quantum computer \cite{Cerezo2021}. On a much shorter term, we foresee experimental setups based on the current scheme; that are readily accessible and able to show the benefits in practice.

\acknowledgments{ We were supported by the Singaporean Ministry Education, via the Tier 2 Grant No. MOE-T2EP50121-0020. \wv{ We thank Micha{\l} Matuszewskia,  Peter McMahon, Fabian Bohm and Thomas van Vaerenbergh for encouraging discussions and comments.}
}


%

\end{document}